\documentclass{article}

\usepackage{arxiv}
\usepackage[utf8]{inputenc}
\usepackage[T1]{fontenc}
\usepackage{hyperref}
\usepackage{orcidlink}
\usepackage{url}
\usepackage{amsmath}
\usepackage{amsfonts}
\usepackage{booktabs}
\usepackage{nicefrac}
\usepackage{microtype}
\usepackage{graphicx}
\usepackage[authoryear, round]{natbib}
\setcitestyle{semicolon,aysep={},yysep={,}}

\usepackage{tikz}
\usepackage{subcaption}
\usepackage{multirow}
\usepackage{placeins}
\usetikzlibrary{quantikz2}

\title{Hybrid quantum learning for imbalanced blood-cell classification: domain adaptation and classifier design}

\author{
  \textbf{Gabriele D'Angeli}\textsuperscript{1,2,}\thanks{Corresponding author: Gabriele D'Angeli. Email: \href{mailto:gabriele.dangeli@hesge.ch}{gabriele.dangeli@hesge.ch}, \href{mailto:gabriele.dangeli@etu.unige.ch}{gabriele.dangeli@etu.unige.ch}.}\,~\orcidlink{0009-0000-8539-5205}\quad
  \textbf{Cyrill Aebischer}\textsuperscript{3}\quad
  \textbf{Eryk Schiller}\textsuperscript{2}\,
  \orcidlink{0000-0003-2187-0382}\quad
  \textbf{Gianluca Nicchiotti}\textsuperscript{3}\,
  \orcidlink{0000-0002-3190-6708}\\[0.4em]
  \textbf{Su Fong Chien}\textsuperscript{4}\,
  \orcidlink{0000-0001-7637-9158}\quad
  \textbf{J{\'e}r{\^o}me Extermann}\textsuperscript{2}\,
  \orcidlink{0000-0002-5858-1122}\quad
  \textbf{Enrico Pomarico}\textsuperscript{2}\,
  \orcidlink{0000-0001-5946-5310}\\[1em]
  \textsuperscript{1}Department of Physics, University of Geneva\\
  Geneva, Switzerland\\[0.5em]
  \textsuperscript{2}Geneva School of Engineering, Architecture
  and Landscape (HEPIA)\\
  HES-SO University of Applied Sciences and Arts Western Switzerland\\
  Geneva, Switzerland\\[0.5em]
  \textsuperscript{3}School of Engineering and Architecture
  of Fribourg (HEIA-FR)\\
  HES-SO University of Applied Sciences and Arts Western Switzerland\\
  Fribourg, Switzerland\\[0.5em]
  \textsuperscript{4}Axon Logic\\
  Multidisciplinary Research and Innovation Centre\\
  Athens, Greece
}

\date{}
\renewcommand{\shorttitle}{Hybrid quantum learning for blood-cell classification}
\hypersetup{
  pdftitle={Hybrid quantum learning for imbalanced blood-cell classification: domain adaptation and classifier design},
  pdfsubject={q-bio.QM, quant-ph, cs.LG},
  pdfauthor={Gabriele D'Angeli},
  pdfkeywords={quantum machine learning, variational quantum classifier, microscopy, leukocyte classification, class imbalance, cross-validation},
  hidelinks
}

\begin{document}

\maketitle

\begin{abstract}
Limited annotated data, class imbalance, and morphological similarity make white blood cell classification challenging. We benchmark hybrid variational quantum classifier heads against an affine baseline on 821 microscopy images across seven morphological classes. Using ImageNet-initialised ResNet34 backbones and repeated stratified five-fold cross-validation, we first compare direct transfer with intermediate adaptation on MLL23, a larger peripheral blood cell microscopy dataset. To guide the head comparison, we explore different encoding strategies, variational depth, and the number of independent circuit blocks. We then compare the classifier performance using the affine head and two hybrid heads across multiple random seeds, under a shared training and evaluation protocol. For this comparison, all backbones share the same MLL23-adapted initialisation. Intermediate adaptation increases mean macro-F1 across all four affine-head fine-tuning configurations. The largest gain occurs with a frozen backbone, where macro-F1 rises from $0.685$ to $0.786$, a relative improvement of approximately $14.7\%$. The shallow two-block hybrid head achieves a mean pooled out-of-fold macro-F1 of $0.865 \pm 0.010$, compared with $0.879 \pm 0.008$ for the affine head. The corresponding balanced accuracies are $0.886 \pm 0.008$ and $0.889 \pm 0.010$, respectively. In the architectural exploration, the two-block hybrid head achieves the highest macro-F1 and balanced accuracy among the evaluated hybrid configurations, while the effect of encoding varies across the considered metrics. Our results help guide the design and evaluation of hybrid classifier heads for imbalanced microscopy by linking architectural choices to overall performance and class-specific precision-recall trade-offs.
\end{abstract}

\keywords{quantum machine learning \and variational quantum classifier \and hybrid quantum-classical learning \and transfer learning \and white blood cell classification}

\section{Introduction}
\label{sec:introduction}
Image classification is challenging when labelled examples are scarce, classes are unevenly represented, and visual differences between categories are subtle. White blood cell (WBC) classification provides a concrete example: morphologically similar categories can have very different numbers of available training samples. Recent research in haematological image analysis has examined class imbalance as an important factor affecting recognition across cell categories \citep{umer2023imbalanceddomaingeneralizationrobust}. In this setting, high overall classification accuracy can coexist with poor performance on under-represented classes. Meaningful evaluation therefore requires both aggregate metrics and class-specific analysis.

Transfer learning helps address the scarcity of labelled images by reusing representations learnt from larger datasets. In haematological microscopy, \citet{koch2024dinobloomfoundationmodelgeneralizable} showed that models trained on cell images learn features that generalise across datasets. Class imbalance can also be addressed during training through class-weighted losses, resampling according to class frequency, and more frequent augmentation of under-represented classes. In these pipelines, the backbone extracts image features and the classifier head maps them to class predictions. Comparing heads can therefore reveal whether alternative ways of processing these features improve recognition of under-represented cell categories.

Quantum machine learning (QML) offers alternative approaches to classifier design. One motivation for studying quantum models with limited training data comes from theoretical work on generalisation. \citet{Caro_2022} established upper bounds on the generalisation error that depend on the size of the training set and the number of trainable circuit gates. The performance of these models relative to classical alternatives remains an empirical question. Hybrid quantum-classical transfer learning provides a practical framework for such comparisons in image classification. In the approach of \citet{Mari_2020}, a pretrained classical network extracts image features, which are then processed by a variational quantum circuit (VQC) and a classical readout.

Previous studies have explored quantum and hybrid models in medical imaging, with some reporting improvements over classical models and others reporting comparable or lower performance. \citet{matic2022quantumclassicalconvolutionalneuralnetworks} reported performance comparable to that of classical convolutional neural networks for breast ultrasound and abdominal CT images in simulation. Similar results were obtained for chest radiographs and retinal images in simulation, although performance declined in some configurations when both training and inference were performed on quantum hardware \citep{Landman_2022}. Hybrid models underperformed their classical counterparts in imbalanced multilabel chest X-ray classification \citep{chan2024expandinghorizonenablinghybrid}, while improvements in macro-F1 were observed in blood-cell microscopy, including with smaller training sets \citep{cruz2026enhancingbloodcellsclassification}.

A systematic review of QML for digital health found no consistent performance advantage over classical methods and highlighted the need for more rigorous comparisons \citep{Gupta_2025}. Such comparisons need to account for both the choice of classical baselines and the role of trainable classical components, since gains cannot automatically be attributed to the quantum circuit \citep{bowles2024betterclassicalsubtleart,sunkel2024quantumimpact}. These considerations motivate evaluating hybrid heads in the context of the complete learning pipeline.

We use MicroscopyRGB, a dataset of 821 peripheral blood cell images derived from Raw-Microscopy \citep{oala2023datamodelsdatasetdrift,oala2021rawmicroscopy}. We compare direct ImageNet transfer with intermediate adaptation on MLL23, a larger peripheral blood cell microscopy dataset. We also evaluate affine and hybrid classifier heads under a shared training and evaluation protocol. An exploration of encoding and circuit architecture guides the head comparison. Whereas \citet{cruz2026enhancingbloodcellsclassification} compared intermediate transformations, we evaluate the replacement of the complete classifier head.

This study therefore addresses three research questions:

\begin{enumerate}
    \item[\textbf{Q1}] Does intermediate adaptation on MLL23 improve target-task performance compared with direct ImageNet transfer, and how does its effect depend on the extent of backbone fine-tuning?

    \item[\textbf{Q2}] How do hybrid VQC heads compare with an affine head in terms of aggregate performance and recognition of under-represented classes under matched training and evaluation conditions?

    \item[\textbf{Q3}] How do encoding strategy, the number of variational layers, and the number of independent circuit blocks relate to aggregate performance and class-specific precision-recall trade-offs?
\end{enumerate}

This study jointly assesses domain adaptation and classifier-head design under a shared protocol for microscopy image classification, considering both aggregate performance and class-specific errors.

\section{Methods}
\label{sec:methods}
The experimental workflow comprises three stages. First, we compare direct ImageNet transfer and intermediate MLL23 adaptation across four backbone fine-tuning configurations using affine heads and three random seeds. The MLL23-adapted initialisation with fine-tuning of layers 3-4 is selected for the subsequent classifier-head comparison. Second, the seven hybrid configurations are explored using one seed under the same target-task training and evaluation protocol. Based on aggregate performance in this exploration, we select the two-block hybrid head for repeated evaluation. In the third stage, this model and a shallow single-block reference are evaluated using two additional seeds. Both configurations use 12 qubits and one variational layer per block. The final comparison therefore includes the affine baseline and these two hybrid heads, each evaluated across three seeds.

\subsection{Data and transfer-learning setup}

\subsubsection{Target dataset}
The Raw-Microscopy dataset contains WBC images acquired in our laboratory from peripheral blood smear slides and labelled by a trained haematological cytologist, as described in the original dataset publication \citep{oala2023datamodelsdatasetdrift,oala2021rawmicroscopy}. For this study, we reuse a single pre-generated RGB representation with $256 \times 256$-pixel images, hereafter referred to as MicroscopyRGB.

Of the original $940$ images, $30$ labelled as unclassified (UNC) are excluded because they do not represent a defined target category. The eight band-neutrophil images (NGB) are also excluded because this class provides very limited support for training and stratified evaluation. We then manually inspect the dataset image by image and remove 81 duplicate images, leaving a final dataset of 821 images across seven morphological categories: basophils, eosinophils, smudge cells, atypical lymphocytes, typical lymphocytes, monocytes, and segmented neutrophils (Table~\ref{tab:class_distribution}).

Images are loaded as $256 \times 256 \times 3$ TIFF files and rescaled to $[0,1]$. Each colour channel is then standardised using the mean pixel intensity and standard deviation for that channel: $\mathbf{\mu} = (0.5566, 0.4607, 0.5923)$ and $\mathbf{\sigma} = (0.0983, 0.1417, 0.0667)$.

\begin{table}[t]
    \caption{Class distribution of the final MicroscopyRGB dataset ($N=821$).}
    \label{tab:class_distribution}
    \centering
    \begin{tabular}{llr}
        \toprule
        Class label & Cell type             & Samples   \\
        \midrule
        BAS         & Basophil              & 18        \\
        EOS         & Eosinophil            & 43        \\
        KSC         & Smudge cell           & 158       \\
        LYA         & Atypical lymphocyte   & 30        \\
        LYT         & Typical lymphocyte    & 224       \\
        MON         & Monocyte              & 180       \\
        NGS         & Segmented neutrophil  & 168       \\
        \midrule
        Total       &                       & 821       \\
        \bottomrule
    \end{tabular}
\end{table}

\subsubsection{Transfer-learning strategies}

The MLL23 dataset contains 41,621 images of peripheral blood cells across 18 morphological classes \citep{ShetabBoushehri2025,shetabboushehri2023mll23}. Images are acquired by automated microscopy at the Munich Leukemia Laboratory and released as $288 \times 288$-pixel TIFF files. Each image is annotated by one of five cytomorphology experts. Several morphological categories overlap with those in MicroscopyRGB, motivating its use for intermediate adaptation.

We compare two transfer-learning paths, both starting from an ImageNet-pretrained ResNet34 \citep{Deng2019ImageNet,russakovsky2015imagenetlargescalevisual}. Direct transfer uses the ImageNet weights to initialise the backbone for training on MicroscopyRGB. The second path includes intermediate adaptation on MLL23 before training on MicroscopyRGB.

For intermediate adaptation, the ImageNet-pretrained ResNet34 is fine-tuned on MLL23 using a stratified $80/20$ training-validation split. Training uses focal loss with $\gamma = 0.5$, label smoothing of $0.05$, and effective-number class weighting. The checkpoint with the highest validation macro-F1 is selected. Its backbone weights are then used to initialise the target-task model, which is trained on MicroscopyRGB with a newly initialised classifier head.

\subsection{Affine and hybrid classification pipelines}

\subsubsection{Feature extraction and affine baseline}

Both pipelines use ResNet34 to extract a compact image representation \citep{he2015deepresiduallearningimage}. After the final residual stage and global average pooling, the backbone maps an input image $\mathbf{I}$ to the $512$-dimensional feature vector 

\begin{equation}
    \mathbf{h} = f_{\phi}(\mathbf{I}) \in \mathbb{R}^{512},
\label{eq:resnet-representation}
\end{equation}

where $\phi$ denotes the parameters of the ResNet34 backbone.

The transfer-learning experiments use either ImageNet or MLL23-adapted initialisation. For the affine-hybrid comparison, all backbones start from the same MLL23-adapted weights. Residual layers 3-4 are fine-tuned jointly with each head, while the stem and residual layers 1-2 remain frozen. This allows the backbone features to adapt to each classifier head.

\begin{figure}[tbp]
    \centering
    \includegraphics[width=\linewidth]{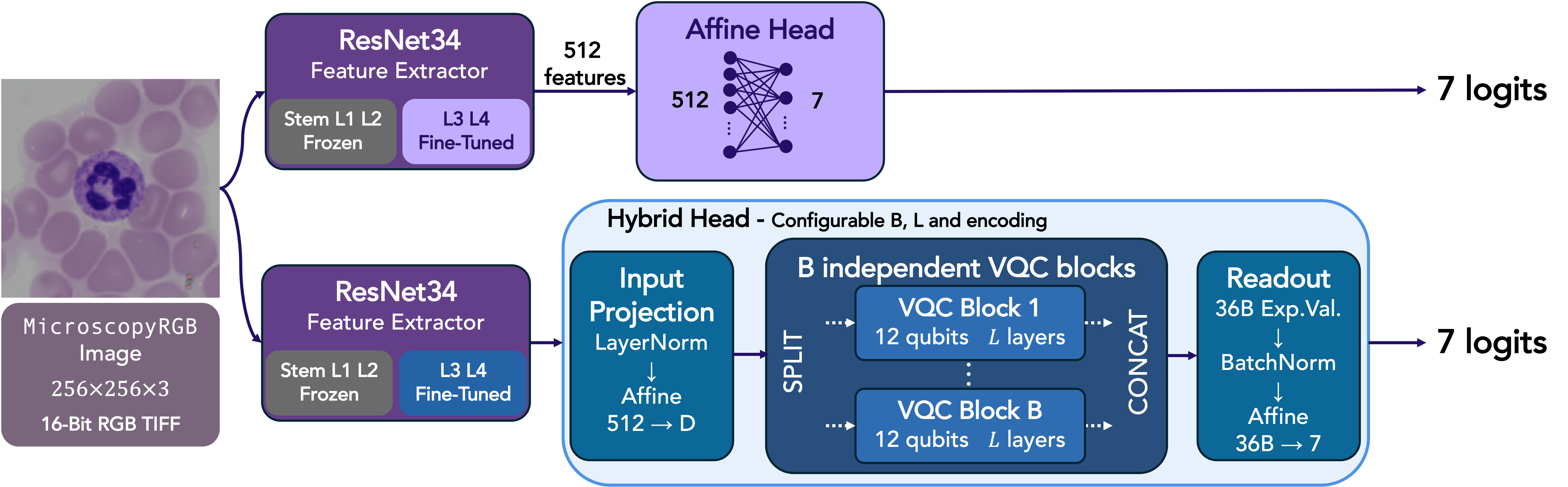}
    \caption{Affine and hybrid classification pipelines used for the head comparison. Both ResNet34 backbones start from the same MLL23-adapted checkpoint and are fine-tuned separately. The hybrid head uses $B$ independent 12-qubit blocks, each with $L$ variational layers. The projection dimension is $D=36B$ for standard re-uploading and $D=36BL$ for incremental re-uploading. Concatenation yields $36B$ expectation values for the classical readout. Input preprocessing and the tanh bounding in Eq.~\ref{eq:angle-map} are omitted}
    \label{fig:classification-pipelines}
\end{figure}

Figure~\ref{fig:classification-pipelines} provides an overview of the affine and hybrid architectures used for the head comparison.

The affine head replaces the original ImageNet classification layer with a newly initialised affine map from the backbone features to the seven target logits, defined by

\begin{equation}
    \mathbf{z}_{\mathrm{C}} = W_{\mathrm{C}}\mathbf{h} +\mathbf{b}_{\mathrm{C}},
\label{eq:classical-head}
\end{equation}

where $W_{\mathrm{C}}\in\mathbb{R}^{7\times512}$ is the weight matrix, $\mathbf{b}_{\mathrm{C}}\in\mathbb{R}^{7}$ is the trainable bias vector, and $\mathbf{z}_{\mathrm{C}}\in\mathbb{R}^{7}$ contains one logit for each class.

\subsubsection{Hybrid head and input encoding}

The hybrid head comprises a classical projection to bounded circuit inputs, one or more quantum circuit blocks, and a classical readout. Encoding influences the functions that the circuit can represent. \citet{Schuld_2021} characterise quantum models with common data-encoding gates through a Fourier representation, showing how encoding operations determine the accessible frequencies and how repeated encoding can enlarge this spectrum. Data re-uploading alternates input-dependent gates with trainable transformations \citep{P_rez_Salinas_2020}. These studies motivate the comparison of encoding strategies described below.

We compare standard re-uploading, which reuses the same projected features at successive layers, with incremental re-uploading, which supplies a different feature slice to each layer. The latter follows the principle introduced by \citet{Periyasamy_2022}, adapted here to features learnt by a classical backbone and projection rather than to a fully quantum classifier.

Let $B$ denote the number of independent circuit blocks, each containing $q_b=12$ qubits, and let $q=Bq_b$ denote their aggregate qubit count. Let $L$ denote the number of variational layers per block. The projection dimension is $D=3q$ for standard re-uploading and $D=3qL$ for incremental re-uploading.
 
The ResNet34 feature vector is layer-normalised, projected by a trainable affine map, and transformed element-wise using a scaled $\mathrm{tanh}$:

\begin{equation}
    \mathbf{x} = \frac{3\pi}{4}\tanh\left[W_{\mathrm{A}}\operatorname{LN}(\mathbf{h})+\mathbf{b}_{\mathrm{A}}\right],
\label{eq:angle-map}
\end{equation}

where LN denotes layer normalisation. The matrix $W_A\in\mathbb R^{D\times512}$ and the vector $b_A\in\mathbb R^D$ are the trainable projection weights and bias, and the subscript A denotes the angle-input projection. The scaled $\mathrm{tanh}$ bounds each component of $\mathbf x$ to the interval $(-3\pi/4,3\pi/4)$.

With standard re-uploading, the same $3q$-dimensional input vector is supplied to every layer:

\begin{equation}
    \boldsymbol{\xi}_l = \mathbf{x}, \qquad l=1,\ldots,L.
\label{eq:standard-input-map}
\end{equation}

With incremental re-uploading, the $3qL$-dimensional projection is split into $L$ layer-specific input vectors, each of dimension $3q$:

\begin{equation}
    \mathbf{x}=
    \begin{bmatrix}
        \boldsymbol{\xi}_1^\top &
        \cdots &
        \boldsymbol{\xi}_L^\top
    \end{bmatrix}^{\top},
    \qquad
    \boldsymbol{\xi}_l\in\mathbb{R}^{3q}.
\label{eq:incremental-input-map}
\end{equation}

For $B>1$, each $\boldsymbol{\xi}_l$ is further partitioned into $B$ contiguous block-specific vectors $\boldsymbol{\xi}_l^{(b)}\in\mathbb{R}^{3q_b}$. We denote by $\xi_{li}^{a,(b)}$ the component assigned to rotation axis $a\in\{X,Y,Z\}$ of qubit $i$ in block $b$ at layer $l$.

\subsubsection{Circuit structure and readout}

Multiple circuits of limited width have been used in hybrid image-classification architectures \citep{Senokosov_2024}. In our implementation, each block contains $12$ qubits and is simulated independently, with no quantum operations connecting different blocks. This design allows the number of encoded features and readout values to increase while keeping the width of each simulated circuit fixed. All trainable components are optimised jointly within each hybrid pipeline.

The joint state of the qubits in block $b$ is initialised as

\begin{equation}
    \lvert \psi_0^{(b)} \rangle = \lvert 0 \rangle^{\otimes q_b}.
\label{eq:block-initial-state}
\end{equation}

At circuit layer $l$, the rotation angle associated with axis $a \in \{X,Y,Z\}$, qubit $i$, and block $b$ is defined as

\begin{equation}
    \gamma_{li}^{a,(b)} = \theta_{li}^{a,(b)} + \omega_{li}^{a,(b)} \xi_{li}^{a,(b)}, \qquad a \in \{X,Y,Z\},
\label{eq:encoded-angle}
\end{equation}

where $\theta_{li}^{a,(b)}$ is a trainable rotation offset and $\omega_{li}^{a,(b)}$ is a trainable encoding scale.

The local single-qubit transformations at layer $l$ are collected into the operator

\begin{equation}
    U_{\mathrm{rot}}^{(l,b)} = \bigotimes_{i=1}^{q_b} \left[ R_Y\left(\gamma_{li}^{Y,(b)}\right)R_Z\left(\gamma_{li}^{Z,(b)}\right) R_X\left(\gamma_{li}^{X,(b)}\right)\right].
\label{eq:rotation-order}
\end{equation}

With the standard right-to-left convention for operator composition, $R_X$, $R_Z$, and $R_Y$ are therefore applied sequentially to each qubit.

Each variational layer applies these local rotations followed by a nearest-neighbour ring of trainable Ising-ZZ interactions, using a hardware-efficient design \citep{Kandala_2017}:

\begin{equation}
    U_{\mathrm{ZZ}}^{(l,b)} = \prod_{(i,j)\in\mathcal{E}_b} \exp\left(-\frac{\mathrm{i}}{2} \varphi_{ij}^{(l,b)} Z_i Z_j \right),
\label{eq:ising-ring}
\end{equation}

where $\varphi_{ij}^{(l,b)}$ is the trainable Ising-$ZZ$ interaction angle for edge $(i,j)$ in block $b$ at layer $l$. The edge set $\mathcal{E}_b$ contains the nearest-neighbour edges of the $b^{\text{th}}$ block, including the closing edge connecting the last qubit to the first.

The state of block $b$ therefore evolves according to

\begin{equation}
    \lvert \psi_l^{(b)} \rangle = U_{\mathrm{ZZ}}^{(l,b)} U_{\mathrm{rot}}^{(l,b)} \lvert \psi_{l-1}^{(b)} \rangle, \qquad l = 1,\ldots,L.
\label{eq:block-layer-evolution}
\end{equation}

After the final layer $L$, a trainable $\operatorname{Rot}(\alpha_i^{(b)},\beta_i^{(b)},\delta_i^{(b)})$ gate is applied to each qubit before readout:

\begin{equation}
    U_{\mathrm{meas}}^{(b)} = \bigotimes_{i=1}^{q_b} \operatorname{Rot}(\alpha_i^{(b)},\beta_i^{(b)},\delta_i^{(b)}), \qquad \lvert \tilde \psi_L^{(b)} \rangle = U_{\mathrm{meas}}^{(b)} \lvert \psi_{L}^{(b)} \rangle
\label{eq:final-rot-ev}
\end{equation}
 
The local Bloch-vector readout of block $b$, evaluated on $\lvert\tilde{\psi}_L^{(b)}\rangle$, is

\begin{equation}
    \begin{split}
        \mathbf{r}^{(b)}
        = ( &\langle X_1\rangle,\ldots,\langle X_{q_b}\rangle,\\
            &\langle Y_1\rangle,\ldots,\langle Y_{q_b}\rangle,\\
            &\langle Z_1\rangle,\ldots,\langle Z_{q_b}\rangle
          )
        \in\mathbb{R}^{3q_b}.
    \end{split}
\label{eq:bloch-readout}
\end{equation}

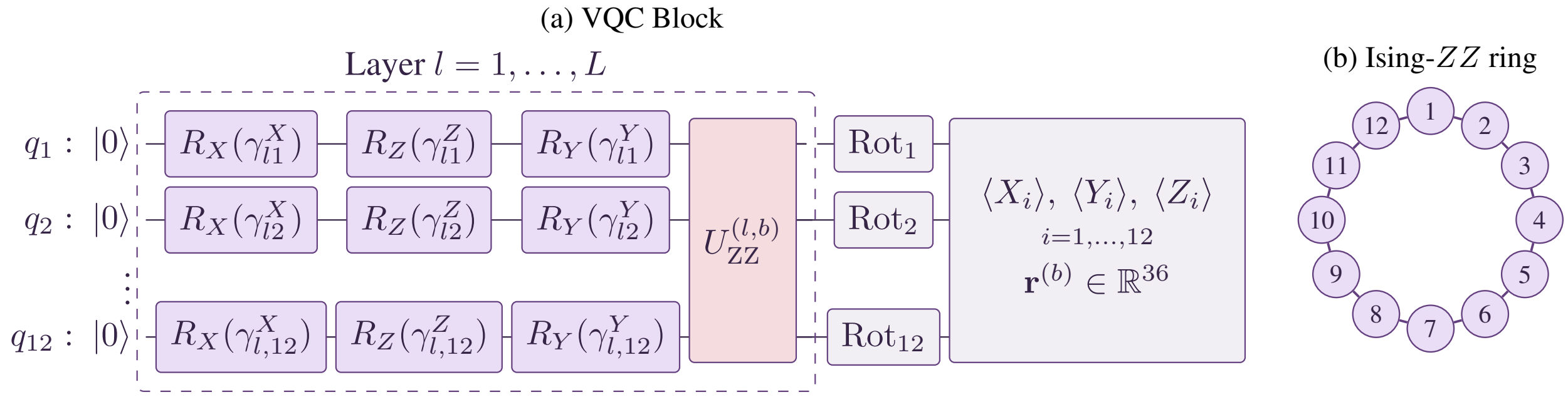
\begin{figure}[tbp]
    \centering
    \begingroup
    \definecolor{vqcInk}{HTML}{362447}
    \definecolor{vqcPurple}{HTML}{654080}
    \definecolor{vqcLilac}{HTML}{E9DEF5}
    \definecolor{vqcRose}{HTML}{F5DDDF}
    \definecolor{vqcGrey}{HTML}{F2EFF4}
    \tikzset{operator/.append style={draw=vqcPurple,fill=vqcLilac,rounded corners=1.5pt,minimum height=0.48cm,inner xsep=3pt,inner ysep=2pt}}
    \captionsetup[subfigure]{position=top,font=small,skip=3pt, justification=centering,singlelinecheck=false}
    
    \begin{subfigure}[c]{0.81\linewidth}
        \centering
        \caption{VQC Block}
        \label{fig:vqc-circuit}
        \resizebox{\linewidth}{!}{%
        \begin{quantikz}[
          row sep={0.78cm,between origins},
          column sep=0.10cm,
          wire types={q,q,n,q},
          thin lines,
          color=vqcInk
        ]
        \lstick{$q_1:\ \lvert0\rangle$}
         & \gate{R_X(\gamma_{l1}^{X})}
           \gategroup[wires=4,steps=4,
             style={draw=vqcPurple,dashed,rounded corners=2pt,
               inner xsep=2pt,inner ysep=2pt},
             label style={label position=above,anchor=south,yshift=-0.15cm,
               inner sep=1pt,text=vqcInk}]
             {Layer $l=1,\ldots,L$}
         & \gate{R_Z(\gamma_{l1}^{Z})}
         & \gate{R_Y(\gamma_{l1}^{Y})}
         & \gate[4,disable auto height,style={fill=vqcRose}]{U_{\mathrm{ZZ}}^{(l,b)}}
         & \qw\hspace{0.12cm}
         & \gate[style={fill=vqcGrey}]{\operatorname{Rot}_1}
         & \gate[4,disable auto height,style={fill=vqcGrey}]{\begin{array}{c}
             \langle X_i\rangle,\ \langle Y_i\rangle,\ \langle Z_i\rangle\\[1pt]
             \scriptstyle i=1,\ldots,12\\[2pt]
             \mathbf r^{(b)}\in\mathbb R^{36}
           \end{array}}
         & \setwiretype{n} \\
        \lstick{$q_2:\ \lvert0\rangle$}
         & \gate{R_X(\gamma_{l2}^{X})}
         & \gate{R_Z(\gamma_{l2}^{Z})}
         & \gate{R_Y(\gamma_{l2}^{Y})}
         &
         & \qw
         & \gate[style={fill=vqcGrey}]{\operatorname{Rot}_2}
         & & \setwiretype{n} \\[-0.18cm]
        \lstick{$\vdots$}
         & & & & & & & & \\[-0.18cm]
        \lstick{$q_{12}:\ \lvert0\rangle$}
         & \gate{R_X(\gamma_{l,12}^{X})}
         & \gate{R_Z(\gamma_{l,12}^{Z})}
         & \gate{R_Y(\gamma_{l,12}^{Y})}
         &
         & \qw
         & \gate[style={fill=vqcGrey}]{\operatorname{Rot}_{12}}
         & & \setwiretype{n}
        \end{quantikz}%
        }
    \end{subfigure}\hfill
    \begin{subfigure}[c]{0.17\linewidth}
        \centering
        \caption{Ising-$ZZ$ ring}
        \label{fig:vqc-ring}
        \resizebox{\linewidth}{!}{%
        \begin{tikzpicture}[
          vqcNode/.style={circle,draw=vqcPurple,fill=vqcLilac,
            text=vqcInk,minimum size=4.5mm,inner sep=0pt,font=\scriptsize},
          vqcEdge/.style={draw=vqcPurple,line width=0.65pt}
        ]
        \foreach \i in {1,...,12}{
          \node[vqcNode] (vqc-\i) at ({90-30*(\i-1)}:1.05cm) {\i};
        }
        \foreach \i [evaluate=\i as \j using int(\i+1)] in {1,...,11}{
          \draw[vqcEdge] (vqc-\i) -- (vqc-\j);
        }
        \draw[vqcEdge] (vqc-12) -- (vqc-1);
        \end{tikzpicture}%
        }
    \end{subfigure}
    \endgroup

    \caption{Structure of one 12-qubit VQC block $b$. (a) Each of the $L$ variational layers applies $R_X$, $R_Z$, and $R_Y$ rotations sequentially, with trainable angle offsets and encoding scales, followed by a trainable Ising-$ZZ$ ring. The final $\operatorname{Rot}_i=\operatorname{Rot}(\alpha_i,\beta_i,\delta_i)$ gates are applied once, after layer $L$. Expectation values of $X$, $Y$, and $Z$ on each qubit form a 36-component readout vector. (b) Nearest-neighbour ring topology, including the 12-1 edge. The block index $b$ is suppressed in local gate labels}
    \label{fig:vqc-block}

\end{figure}

Figure~\ref{fig:vqc-block} summarises the variational layers and Bloch-vector readout of one 12-qubit block.

For multiple blocks, these vectors are concatenated in block order. The resulting vector $\mathbf{r}\in\mathbb{R}^{3q}$ is batch-normalised and mapped to seven logits:

\begin{equation}
    \mathbf{z}_{\mathrm{H}}
    =
    W_{\mathrm{H}}\operatorname{BN}(\mathbf{r})
    +\mathbf{b}_{\mathrm{H}},
    \label{eq:hybrid-head}
\end{equation}

where BN denotes batch normalisation, and the subscript $\mathrm H$ denotes the hybrid head. The readout has trainable weights $W_{\mathrm H}\in\mathbb{R}^{7\times3q}$ and bias $\mathbf b_{\mathrm H}\in\mathbb{R}^{7}$. Its output $\mathbf z_{\mathrm H}\in\mathbb{R}^{7}$ contains one logit for each target class.
 
\subsubsection{Initialisation, parameter counts, and comparison scope}

Encoding scales are initialised to one. Rotation offsets, Ising-$ZZ$ angles, and final measurement-basis rotation parameters are independently drawn from $\mathcal{N}(0,0.01)$. Circuit expectation values are computed analytically using noiseless statevector simulation.
 
Each block is a 12-qubit ring with $|\mathcal{E}_b|=q_b$ edges. The number of trainable circuit parameters is therefore
\begin{equation}
    \begin{aligned}
        N_{\mathrm{circuit}}    &= B\left(6Lq_b+L|\mathcal{E}_b|+3q_b\right)\\
                                &= Bq_b(7L+3)\\
                                &= q(7L+3).
    \end{aligned}
\label{eq:circuit-parameters}
\end{equation}

The first term contains the rotation offsets and encoding scales, the second the Ising-$ZZ$ angles, and the third the final measurement-basis rotations.
 
Table~\ref{tab:pipeline-parameters} reports parameter counts for the shallow configurations. The hybrid heads contain more trainable post-backbone parameters than the affine baseline, although the difference is small relative to the backbone parameter count. The comparison evaluates complete head replacement and therefore does not isolate the contribution of the VQC from that of the additional classical projection, normalisation, and readout.
 
\begin{table}[t]
  \caption{Parameter counts for the shallow MLL23-adapted configurations. Hybrid rows use $L=1$, an affine angle projection with tanh bounding, Bloch-vector readout, and trainable Ising-$ZZ$ and final measurement-basis rotations. Trainable totals assume the layer-3-4 fine-tuning configuration. Circuit parameters are included in the post-backbone head count.}
  \label{tab:pipeline-parameters}
  \centering
  \small
  \begin{tabular}{lrrrrr}
    \toprule
    Model & Max. circuit width & Post-backbone head & Circuit params. & Total & Trainable \\
    \midrule
    ResNet34 + affine head & - & 3,591 & - &
        21,288,263 & 19,940,359 \\
    Hybrid $B1/L1$ & 12 & 19,943 & 120 &
        21,304,615 & 19,956,711 \\
    Hybrid $B2/L1$ & 12 & 38,855 & 240 &
        21,323,527 & 19,975,623 \\
    Hybrid $B3/L1$ & 12 & 57,767 & 360 &
        21,342,439 & 19,994,535 \\
    \bottomrule
  \end{tabular}
\end{table}
 
\subsection{Training and evaluation protocol}
\label{sec:training-evaluation}

The target-task head comparison uses shared fold assignments, backbone initialisation and fine-tuning settings, augmentation, imbalance handling, and optimisation settings. Intermediate adaptation on MLL23 is conducted separately using the procedure described above.

\subsubsection{Cross-validation design}

Target-task performance is assessed using shuffled five-fold cross-validation stratified by class. Each image appears in a validation fold exactly once per seed. No separate test set is allocated, given the limited total sample size and low support of several classes. Performance is therefore estimated from pooled out-of-fold (OOF) predictions.

The transfer-learning and classifier-head comparisons use three randomly chosen seeds: $250$, $9$, and $122$. The architectural exploration uses the first of these, and the two selected hybrid heads are then evaluated with the other two. All compared pipelines use identical fold assignments within each seed. The separate stratified training-validation split for intermediate MLL23 adaptation uses seed $20$.

\subsubsection{Loss and imbalance handling}

For each fold, the model is trained for $50$ epochs with a batch size of $64$. The objective is class-weighted cross-entropy with label smoothing $\epsilon=0.05$. For $C=7$ classes, the smoothed target distribution is

\begin{equation}
    \widetilde{y}_{ic} =  (1-\epsilon)\mathbf{1}[c=y_i] + \frac{\epsilon}{C}.
\label{eq:label-smoothing}
\end{equation}

Class weights are calculated separately from the training partition of each fold using the effective number of observations \citep{cui2019classbalancedlossbasedeffective}. For class $c$ with $n_c$ training images,

\begin{equation}
    E_c = \frac{1-\beta^{n_c}}{1-\beta},\qquad w_c= \frac{C E_c^{-1}} {\sum_{k=1}^{C}E_k^{-1}}, \qquad \beta=0.999.
\label{eq:effective-number-weights}
\end{equation}

The weights have unit mean across classes and are applied within the smoothed cross-entropy loss to increase the contribution of under-represented categories.

Training images are also sampled with replacement using inverse-frequency weights $s_i=1/n_{y_i}$. Because the weights sum to one within each class, all classes have equal expected sampling probability. Each epoch contains as many draws as images in the training fold, allowing an image to be sampled more than once or omitted in a given epoch. The same loss and sampling procedures are used across target-task comparisons.

\subsubsection{Conditional data augmentation}

Augmentation is applied only to training images, before channel-wise normalisation. Eligibility depends on class support in the current training fold. Images from classes with at most 30 training observations, BAS and LYA, are always eligible for augmentation, while images from the remaining classes are eligible with probability $0.3$.

Geometric transformations comprise horizontal and vertical flips, each applied with probability $0.5$, and a randomly selected rotation of $90^\circ$, $180^\circ$, or $270^\circ$, applied with probability $0.5$. Brightness and contrast are adjusted with probability $0.35$. The augmentation pipeline also includes additive Gaussian noise (with probability $0.25$), binomial blur (with probability $0.20$), and sharpness adjustment (with probability $0.35$).

\subsubsection{Optimisation and epoch selection}

All trainable components are optimised jointly using AdamW \citep{loshchilov2019decoupledweightdecayregularization}. For the trainable ResNet34 layers we use a learning rate of $3\times10^{-4}$, while the model-specific post-backbone head uses $10^{-3}$. Weight decay is set to $10^{-3}$. In hybrid models, circuit rotation offsets, encoding scales, entangling angles, and final measurement-basis rotations use the head learning rate with zero weight decay.

A \texttt{ReduceLROnPlateau} scheduler monitors the validation macro-F1 score. The minimum learning rate for each parameter group is set at $1\%$ of its initial value. For each fold, validation predictions from the epoch with the highest macro-F1 are retained for OOF aggregation.

Hybrid pipelines are trained end-to-end through the trainable backbone layers, angle projection, circuits, and classical readout. Circuits are evaluated with the \texttt{lightning.gpu} backend and adjoint differentiation on an NVIDIA A40 GPU.

\subsubsection{Metrics and OOF aggregation}

Macro-F1 is the primary aggregate metric. It assigns equal weight to each of the $C = 7$ classes and is computed from the precision $P_c$ and recall $R_c$ of each class $c$:

\begin{equation}
  F_{1,\mathrm{macro}}
  =
  \frac{1}{C}
  \sum_{c=1}^{C}
  \frac{2P_cR_c}{P_c+R_c},
  \label{eq:macro-f1}
\end{equation}

where undefined class-specific values are set to zero.

Complementary metrics are accuracy, weighted F1, macro precision, and class-specific precision, recall, and F1. Balanced accuracy is calculated as

\begin{equation}
  \operatorname{BA}
  =
  \frac{1}{C}\sum_{c=1}^{C}R_c.
  \label{eq:balanced-accuracy}
\end{equation}

All metrics are computed using scikit-learn \citep{pedregosa2018scikitlearnmachinelearningpython}.

For each configuration and seed, we pool validation predictions from the selected epoch of each fold, yielding one prediction for each of the 821 images. Aggregate and class-specific metrics, as well as the confusion matrix, are calculated directly from these pooled predictions.

For repeated configurations, pooled OOF metrics are calculated separately for each seed and reported as the mean and sample standard deviation across the three seeds.

\subsubsection{Experiment tracking and computational cost}

We track experiments using Weights \& Biases \citep{biewald2020wandb} and record end-to-end training times. Hybrid runtimes are measured under noiseless statevector simulation.

\section{Results}
\label{sec:results}

Results address the research questions introduced in Section~\ref{sec:introduction}. For the transfer-learning and classifier-head comparisons, we report the mean and sample standard deviation of the evaluation metrics across three seeds. The architectural exploration uses a single seed.

\subsection{Intermediate MLL23 adaptation}

\begin{figure}[tbp]
    \centering

    \begin{subfigure}[t]{0.40\linewidth}
        \centering
        \vspace{0pt}
        \phantomcaption
        \label{fig:transfer-macro-f1-a}
        \includegraphics[width=\linewidth]{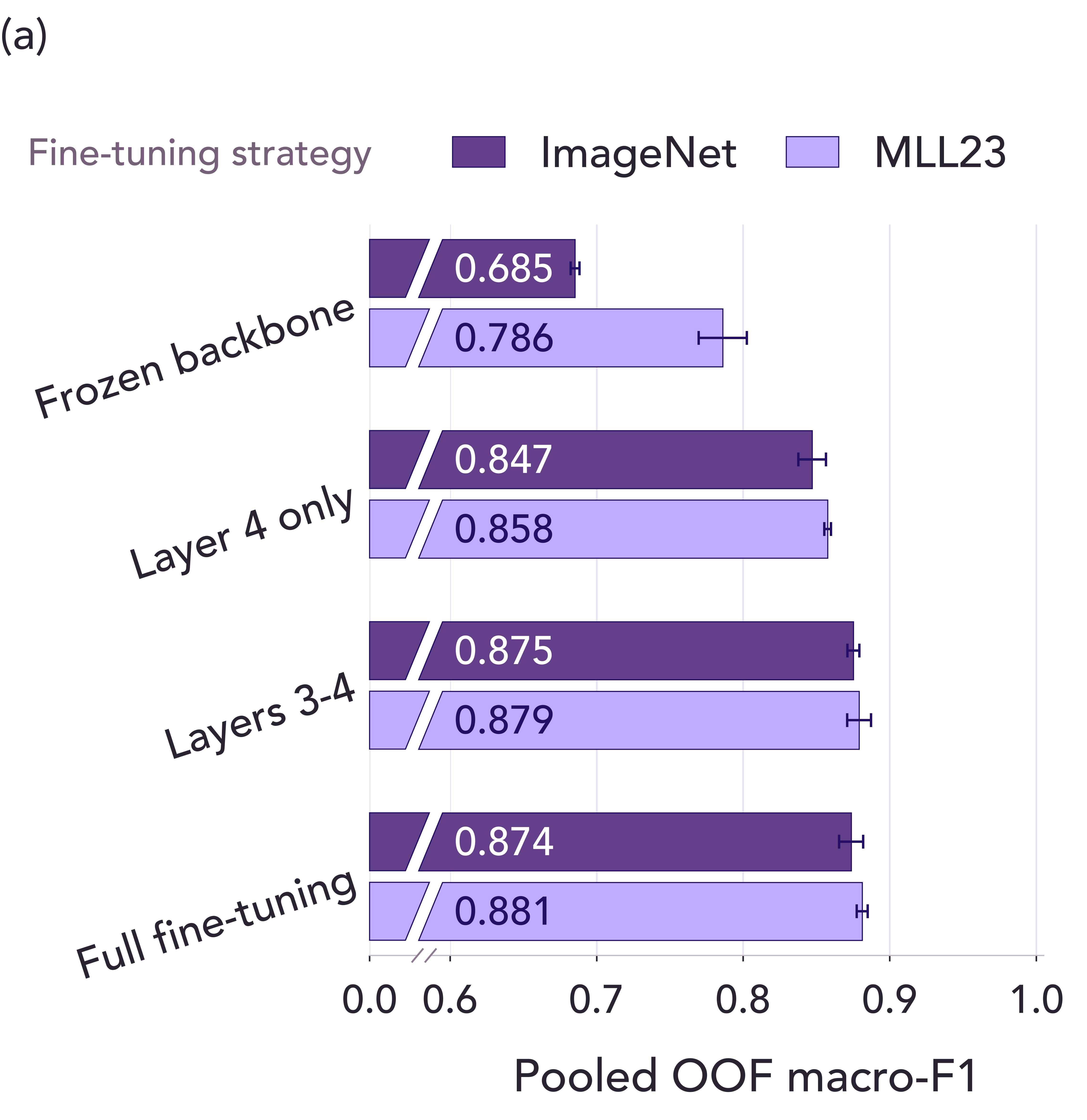}
    \end{subfigure}\hfill
    \begin{subfigure}[t]{0.58\linewidth}
        \centering
        \vspace{0pt}
        \phantomcaption
        \label{fig:transfer-macro-f1-b}
        \includegraphics[width=\linewidth]{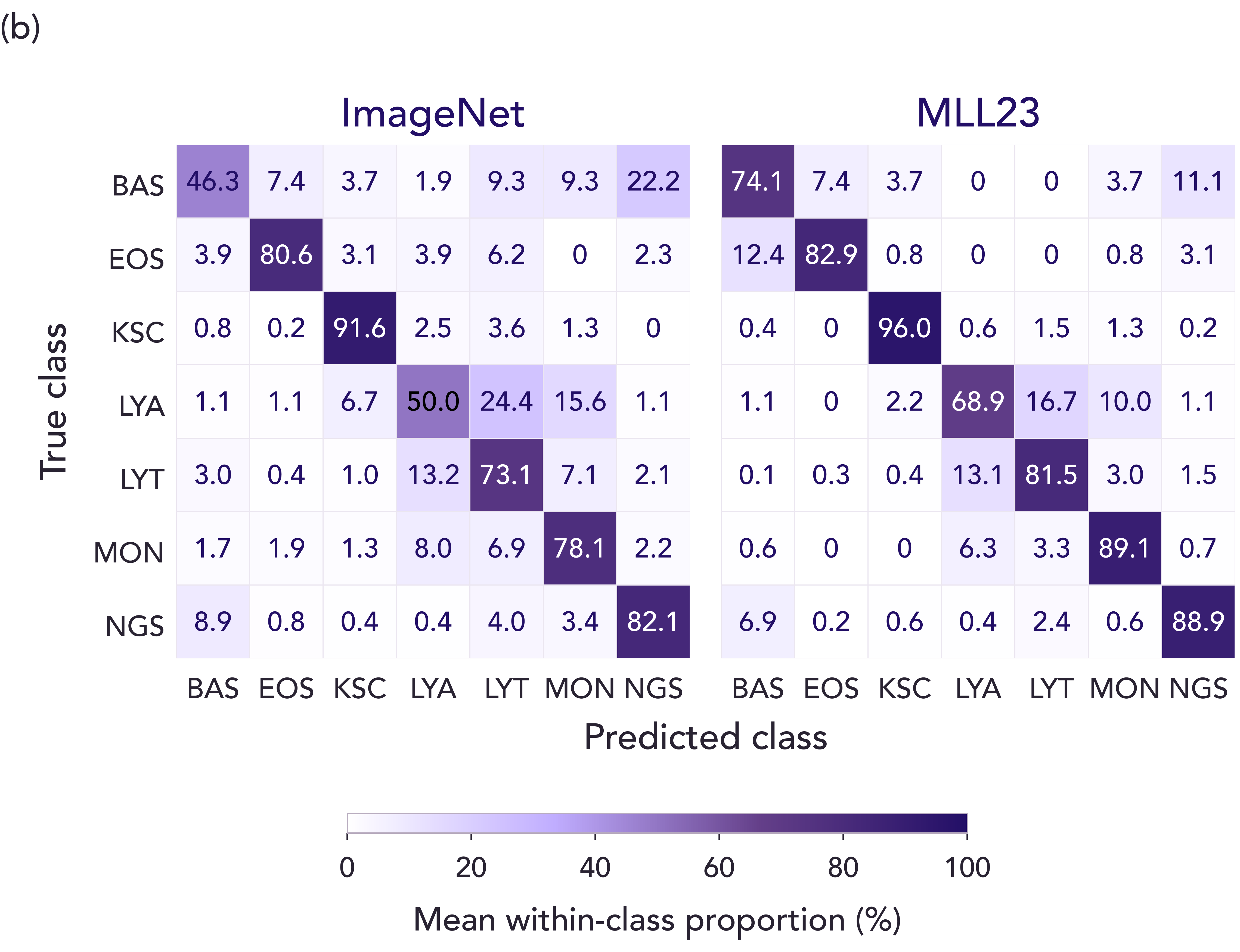}
    \end{subfigure}

    \caption{Macro-F1 scores and confusion matrices with and without intermediate MLL23 adaptation. \textbf{(a)} Mean pooled out-of-fold (OOF) macro-F1 for four ResNet34 fine-tuning configurations. Error bars show sample standard deviations across seeds 9, 122, and 250. \textbf{(b)} Confusion matrices for frozen-backbone models using direct ImageNet transfer and MLL23-adapted initialisation. For each seed, the pooled OOF matrix is normalised by true class, and corresponding entries are then averaged across the three seeds. Cell values and colours indicate mean within-class percentages. Diagonal entries represent mean class recall. All configurations use affine classifier heads and five-fold cross-validation on the same 821 images for each seed}
    \label{fig:transfer-macro-f1}
\end{figure}

Intermediate MLL23 adaptation increases mean pooled OOF macro-F1 in all four affine-head configurations (Figure~\ref{fig:transfer-macro-f1-a}). The largest gain occurs with a frozen backbone, where macro-F1 rises from $0.685$ to $0.786$, a relative improvement of approximately $14.7\%$. With target-task fine-tuning, the increases are smaller, ranging from $0.004$ to $0.011$. Aggregate metrics for all four fine-tuning configurations are reported in Table~\ref{tab:transfer-results-full} in Appendix~\ref{sec:supp-metrics}.

The frozen-backbone confusion matrices show higher mean recall for all seven classes, with the largest improvements for the two least-supported classes (Figure~\ref{fig:transfer-macro-f1-b}). Mean recall increases from $46.3\%$ to $74.1\%$ for basophils (BAS, $n=18$) and from $50.0\%$ to $68.9\%$ for atypical lymphocytes (LYA, $n=30$). Mean precision and F1 also increase for all seven classes (Table~\ref{tab:transfer-class-results}).

With MLL23-adapted initialisation, fine-tuning residual layers 3-4 yields a mean macro-F1 of $0.879$, compared with $0.881$ for full fine-tuning. Given this small difference, we use the layers 3-4 configuration for the head comparison, allowing the later features to adapt to each classifier while keeping the stem and residual layers 1-2 frozen.

\subsection{Affine-hybrid comparison}
Among the seven hybrid configurations evaluated using a single seed, the shallow two-block head $B2/L1$ achieves the highest macro-F1 and balanced accuracy (Table~\ref{tab:hybrid-architecture-results}). We therefore include it in the comparison with the affine head, alongside $B1/L1$ as a single-block reference. Both hybrid heads are also evaluated using two additional seeds.

\begin{figure}[tbp]
    \centering

    \begin{minipage}[t]{0.38\linewidth}
        \centering
        \vspace{0pt}
        \includegraphics[width=\linewidth]{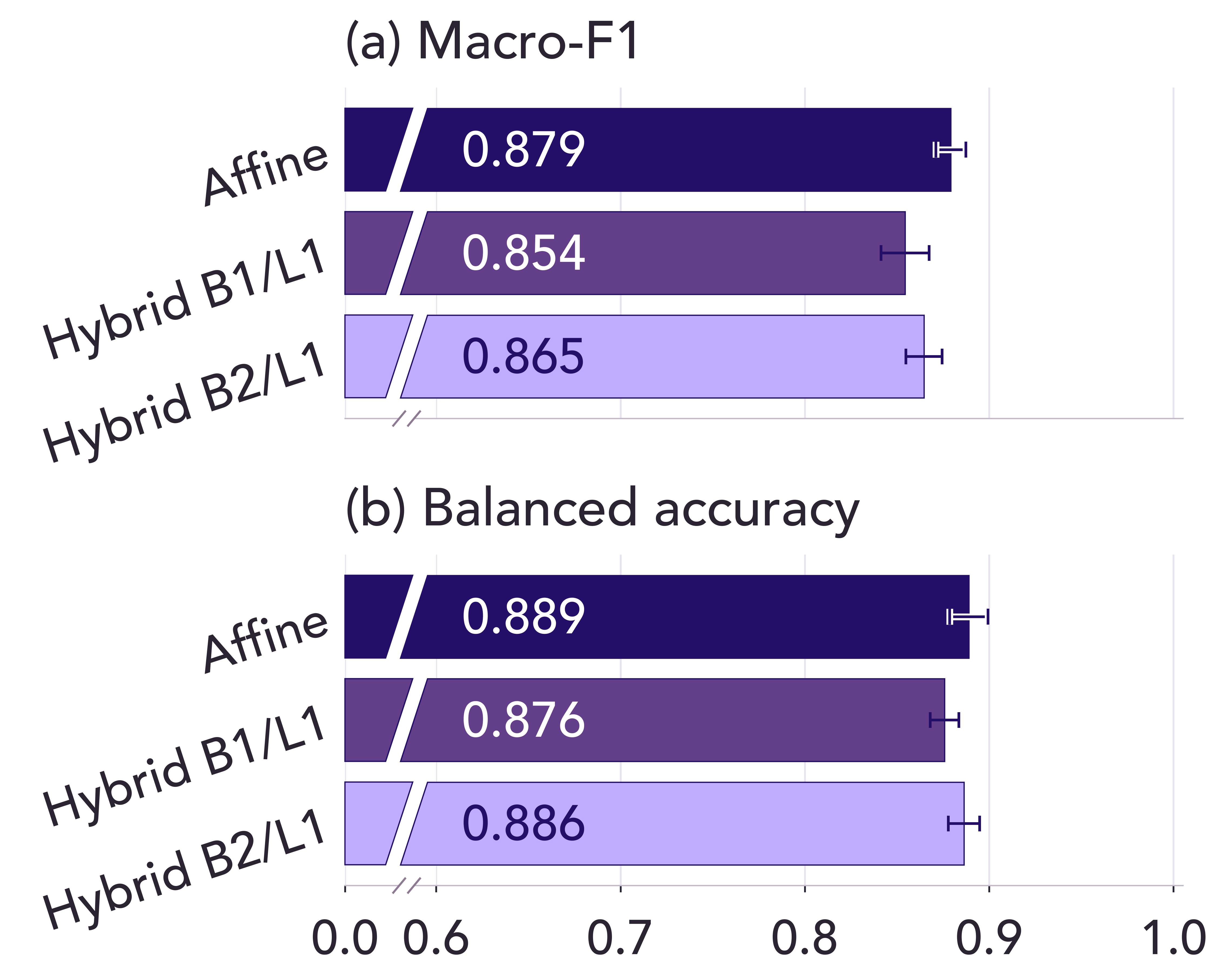}
    \end{minipage}\hfill
    \begin{minipage}[t]{0.60\linewidth}
        \centering
        \vspace{0pt}
        \includegraphics[width=\linewidth]{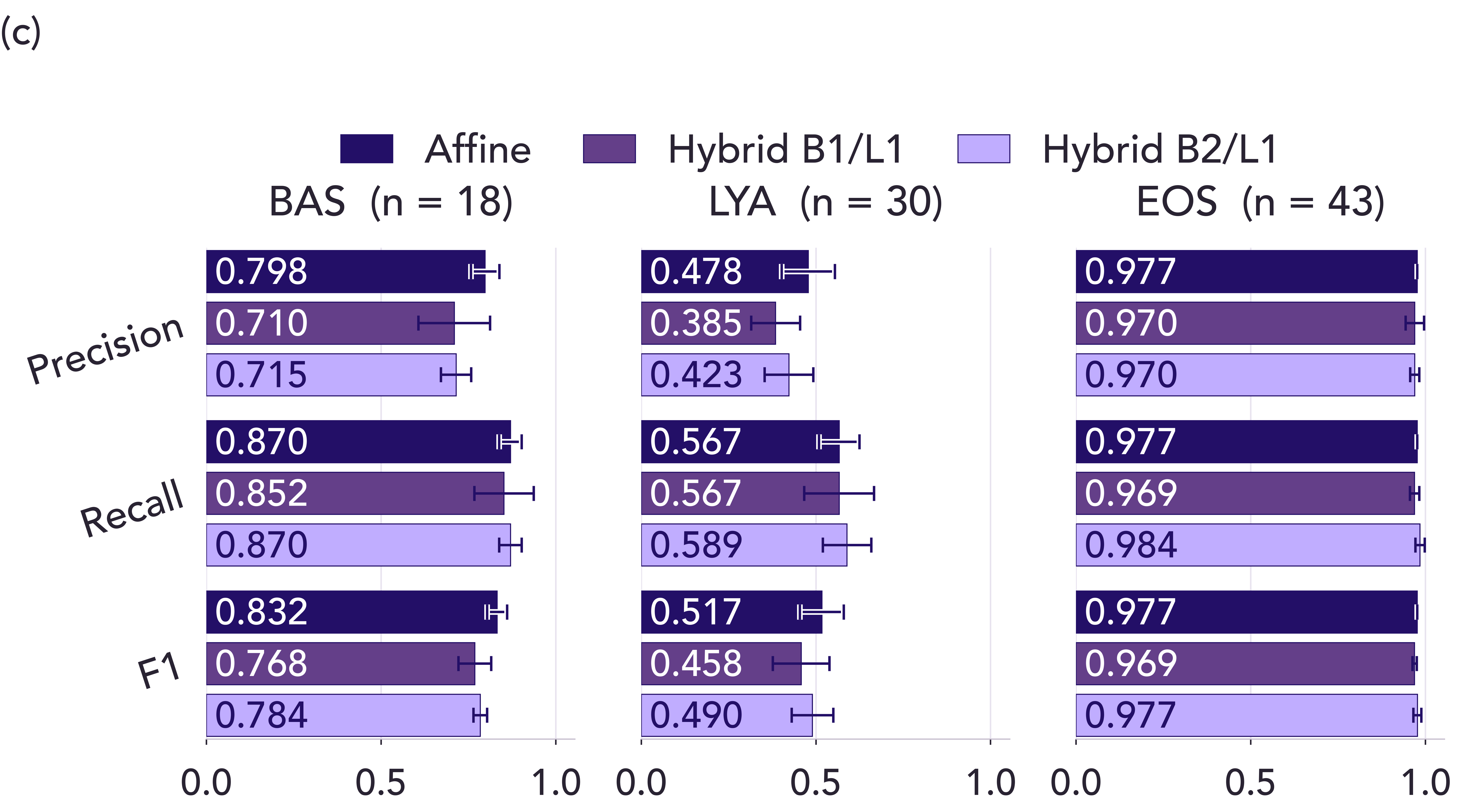}
    \end{minipage}

    \caption{Comparison of the affine head and the two selected hybrid heads. \textbf{(a)} Pooled OOF macro-F1. \textbf{(b)} Balanced accuracy. \textbf{(c)} Class-specific precision, recall, and F1 for BAS ($n=18$), LYA ($n=30$), and EOS ($n=43$). Bar labels show means across seeds 9, 122, and 250, and error bars show standard deviations. Metrics are calculated from pooled OOF predictions separately for each seed. All models start from the same MLL23-adapted ResNet34 weights and fine-tune residual layers 3-4 jointly with the classifier head}
    \label{fig:head-rare-comparison}
\end{figure}

The affine head achieves the highest mean macro-F1 ($0.879 \pm 0.008$), followed by $B2/L1$ ($0.865 \pm 0.010$) and $B1/L1$ ($0.854 \pm 0.013$; Figure~\ref{fig:head-rare-comparison}a). Balanced accuracy is numerically similar for the affine head and $B2/L1$, at $0.889 \pm 0.010$ and $0.886 \pm 0.008$, respectively (Figure~\ref{fig:head-rare-comparison}b). Table~\ref{tab:head-results-full} reports the full aggregate results, including accuracy, macro precision, and weighted F1.

The low-support classes show different precision-recall trade-offs (Figure~\ref{fig:head-rare-comparison}c and Table~\ref{tab:low-support-results}). Compared with the affine head, $B2/L1$ has slightly higher mean recall for LYA ($0.589$ versus $0.567$) and EOS ($0.984$ versus $0.977$), while BAS recall remains at $0.870$. For LYA, higher recall comes with lower precision ($0.423$ versus $0.478$) and F1 ($0.490$ versus $0.517$). BAS F1 is also lower ($0.784$ versus $0.832$), whereas EOS F1 remains approximately $0.977$. $B1/L1$ has the same mean LYA recall as the affine head but lower recall for BAS and EOS. The recall gains observed for $B2/L1$ therefore do not translate into a consistent F1 improvement across these classes.

Seed-specific confusion matrices are provided in Figures \ref{fig:confusion-seed9}-\ref{fig:confusion-seed250}. For the seed used in the architectural exploration, $B2/L1$ correctly classifies 20 of the 30 LYA images, compared with 16 for the affine head, but produces 26 rather than 15 false-positive LYA predictions (Figure \ref{fig:confusion-seed250}).

\subsection{Encoding strategy, variational layer count, and independent blocks}
Table~\ref{tab:hybrid-architecture-results} compares seven hybrid configurations evaluated using a single seed. With single encoding, $B2/L1$ achieves the highest macro-F1 ($0.869$) and balanced accuracy ($0.888$), followed by $B1/L1$ and $B3/L1$. Adding blocks also increases the size of the classical projection and readout.

\begin{table}[t]
    \centering
    \caption{Macro-F1, balanced accuracy, and class-specific precision and recall for seven MLL23-adapted hybrid configurations evaluated using a single seed.}
    \label{tab:hybrid-architecture-results}
    \vspace{4pt}
    \begingroup
    \small
    \renewcommand{\arraystretch}{1.22}
    \setlength{\tabcolsep}{4pt}

    \begin{tabular*}{\linewidth}
        {@{\extracolsep{\fill}}llrrrrrr@{}}
        \toprule

        \multirow{2}{*}{\normalsize\bfseries Encoding}
        & \multirow{2}{*}{\normalsize\bfseries Config.}
        & \multicolumn{2}{c}{\normalsize\bfseries Aggregate metrics}
        & \multicolumn{2}{c}{\normalsize\bfseries BAS}
        & \multicolumn{2}{c}{\normalsize\bfseries LYA} \\

        \cmidrule(lr){3-4}
        \cmidrule(lr){5-6}
        \cmidrule(lr){7-8}

        & &
        Macro-F1
        & Balanced accuracy
        & Precision & Recall
        & Precision & Recall \\
        \midrule

        \multirow{3}{*}{Single encoding}
        & $B1/L1$
        & 0.866 & 0.873
        & 0.824 & 0.778
        & 0.415 & 0.567 \\

        & $B2/L1$
        & 0.869 & 0.888
        & 0.750 & 0.833
        & 0.435 & 0.667 \\

        & $B3/L1$
        & 0.854 & 0.865
        & 0.750 & 0.833
        & 0.385 & 0.500 \\

        \addlinespace[4pt]

        Standard re-uploading
        & $B1/L2$
        & 0.804 & 0.858
        & 0.452 & 0.778
        & 0.292 & 0.700 \\

        \addlinespace[4pt]

        \multirow{3}{*}{Incremental re-uploading}
        & $B1/L2$
        & 0.830 & 0.847
        & 0.667 & 0.667
        & 0.327 & 0.600 \\

        & $B1/L3$
        & 0.800 & 0.833
        & 0.500 & 0.722
        & 0.258 & 0.533 \\

        & $B2/L2$
        & 0.840 & 0.870
        & 0.615 & 0.889
        & 0.340 & 0.533 \\

        \bottomrule
    \end{tabular*}
    \endgroup

    \par\vspace{4pt}

    \begin{minipage}{\linewidth}
        \footnotesize
        \raggedright
        All values are pooled OOF estimates for the same 821 images. Configuration labels specify the number of independent 12-qubit blocks ($B$) and variational layers per block ($L$). Single encoding denotes one input-encoding step per block ($L=1$).
    \end{minipage}
\end{table}

For $B1/L2$, incremental re-uploading yields higher macro-F1 than standard re-uploading ($0.830$ versus $0.804$), but lower balanced accuracy ($0.847$ versus $0.858$). BAS and LYA both show higher precision and lower recall.

For the single-block heads, macro-F1 decreases from $0.866$ with single encoding at $L=1$ to $0.804$ with standard re-uploading at $L=2$. With incremental re-uploading, increasing $L$ from two to three reduces macro-F1 from $0.830$ to $0.800$. Balanced accuracy also decreases in both comparisons.

Under noiseless statevector simulation, $B1/L1$ and $B2/L1$ require approximately $101$ and $198$ times the training time of the affine baseline. Table~\ref{tab:computational-cost} reports runtimes, trainable head parameter counts, and measurement conditions.

\section{Discussion}
\label{sec:discussion}

\paragraph{Dataset adaptation and feature learning} The MLL23 adaptation improves the mean macro-F1 score across all four fine-tuning configurations, with the greatest improvement observed when the backbone is totally frozen. Since only the affine head learns from the target images in this setting, this suggests that the adapted features better distinguish the target classes. Recall improves across all seven classes, particularly BAS and LYA. On the other hand, the smaller gains after fine-tuning may indicate that target-task learning can partly compensate for differences in the initial features.

The comparison between the different heads is based on the hybrid transfer-learning framework described in \citet{Mari_2020}, in which the subsequent layers of the backbone are trained jointly with each head. The sharing of initial weights does not imply that the final features are identical: the results also reflect the way in which the base network adapts to each classifier.

\paragraph{Beyond aggregate performance} The similar balanced accuracy of the affine and two-block heads masks differences in their errors. For $B2/L1$, which uses two independent 12-qubit blocks with one variational layer each, higher mean LYA recall comes with lower precision and F1. Class-specific analysis therefore helps to determine whether an improvement in recall leads to an increase in false positives.

The high F1 score for EOS, despite its limited support, may partly reflect the characteristic cytoplasmic granules of eosinophils \citep{scordino2016eosinophil}. Atypical lymphocytes, by contrast, have a variable appearance and may resemble typical lymphocytes or monocytes \citep{nozaka2023classyfingmicroscopicimages}. This morphological overlap would explain the frequent confusion of LYA with LYT and MON, but makes the classification problem particularly challenging when there are few training samples.

\paragraph{Circuit depth} Repeated encoding can enlarge the frequency spectrum accessible to quantum models \citep{Schuld_2021}, but greater representational capacity does not guarantee better classification. The deeper single-block configurations have lower aggregate scores and longer training times.

\paragraph{Encoding and classical components} The performance of $B2/L1$ in the architectural exploration motivates further study of independent blocks, while the encoding comparison shows that gains depend on the metric considered. These architectural choices also change the classical components: incremental re-uploading enlarges the projection, and additional blocks enlarge both the projection and readout. \citet{cruz2026enhancingbloodcellsclassification} kept the surrounding architecture fixed when comparing intermediate transformations and reported a higher macro-F1 score for hybrid models. Our study evaluates complete head replacement following MLL23 adaptation, with a particular focus on a small, imbalanced reference dataset that distinguishes between typical and atypical lymphocytes.

\paragraph{Limitations} One of the limitations of the evaluation protocol concerns the overlap between model development and performance evaluation. This choice was motivated by the limited number of available samples, particularly in the least-supported classes. Performance on the validation folds was used to adjust learning rates and select the epoch with the best performance, whilst the development of the architecture was based on the same benchmark dataset used for evaluation. The aggregated OOF estimates may therefore be influenced by selection bias~\citep{Varma2006BiasIE, Cawley2010OnOI}.

\section{Conclusion}
\label{sec:conclusion}
This study shows how domain adaptation and classifier-head design affect blood-cell classification on an imbalanced microscopy dataset. Intermediate MLL23 adaptation improves mean macro-F1 across the four affine-head fine-tuning settings, with the largest gain when the backbone is frozen. With the MLL23-adapted backbone, the shallow two-block hybrid head achieves mean macro-F1 and balanced accuracy scores that are numerically comparable to those of the affine head, given the associated uncertainties. The comparison also reveals distinct precision-recall trade-offs across classes.

The architectural exploration favours shallow configurations and shows that encoding and block choices are associated with different patterns of classification errors. Taken together, these results provide practical guidance for selecting hybrid classifiers based on both overall performance and the trade-off between precision and recall for under-represented classes. Future work could extend the hybrid approach by integrating quantum components into feature extraction and by exploring trainable quantum pre-processing modules. Joint optimisation of these components with the classifier could clarify whether quantum pre-processing improves the ability to distinguish between morphologically similar cell types, particularly those for which training data is limited.

\section*{Statements and Declarations}
\paragraph*{Funding}
This work was carried out as part of the SecQDevOps project (Trustworthy Quantum DevOps for Secure Software and Hardware Quantum Engineering), funded by the European Union under the Horizon Europe research and innovation programme (Grant Agreement No.~101225776). This work has received funding from the Swiss State Secretariat for Education, Research and Innovation (SERI) through its support for HES-SO's participation in SecQDevOps (project No.~25.00261).

\paragraph*{Competing interests}
The authors declare no competing interests.

\paragraph*{Author contributions}
G.D., E.P., J.E., E.S., and G.N. conceived the study. G.D. and C.A. developed the methodology and software and conducted the experiments. G.D. curated and analysed the data, prepared the figures, and drafted the manuscript. G.D., G.N., E.S., S.F.C., J.E., and E.P. reviewed and edited the manuscript. J.E. and E.P. supervised the project. All authors approved the manuscript for submission.

\paragraph*{Ethics approval}
This study uses existing images from the Raw-Microscopy and MLL23 datasets. Raw-Microscopy comprises images acquired from commercially obtained, de-identified human blood smear slides \citep{oala2023datamodelsdatasetdrift}. The original MLL23 work received approval from the Ethics Committee of LMU Munich (reference number 25-0744) \citep{ShetabBoushehri2025}. No new biological samples were collected for this study.

\paragraph*{Data and code availability}
The Raw-Microscopy dataset and associated data models are described by \citet{oala2023datamodelsdatasetdrift} and archived on Zenodo \citep{oala2021rawmicroscopy}. The MLL23 dataset is described by \citet{ShetabBoushehri2025} and available on Zenodo \citep{shetabboushehri2023mll23}.

The implementation is available on GitHub at \url{https://github.com/gabrieledangeli/hybrid-quantum-classical-microscopy-classification}. The associated Weights \& Biases project is available at
\url{https://wandb.ai/gabriele-dangeli2-hes-so-secqdevops/MicroscopyRGB_Classification}. The filenames of the 81 duplicate images removed from MicroscopyRGB and the MLL23-adapted ResNet34 backbone weights used in this study, together with checkpoint metadata and loading instructions, are publicly available on Zenodo \citep{dangeli2026microscopyresources}.

\section*{Acknowledgements}
We thank Nicolas Brunner and Pavel Sekatski for stimulating discussions about the quantum information aspects of this work, and Michèle Buck and Christian Matek for helpful discussions on data labelling.

We used ChatGPT (OpenAI) to improve the grammar and style of the manuscript. The authors reviewed and edited all suggested revisions and take full responsibility for the final content.

\clearpage

\bibliographystyle{sn-basic}
\bibliography{references}

\clearpage

\appendix

\appendix

\setcounter{table}{0}
\renewcommand{\thetable}{S\arabic{table}}
\renewcommand{\theHtable}{supp.\arabic{table}}

\setcounter{figure}{0}
\renewcommand{\thefigure}{S\arabic{figure}}
\renewcommand{\theHfigure}{supp.\arabic{figure}}

\raggedbottom
\makeatletter
\setlength{\@fptop}{0pt}
\makeatother

\section{Aggregate and class-specific performance}
\label{sec:supp-metrics}

Tables~\ref{tab:transfer-results-full}-\ref{tab:low-support-results} report aggregate and class-specific results for the transfer-learning and classifier-head comparisons. For each seed ($9$, $122$, and $250$), the evaluation metrics are calculated from pooled five-fold OOF predictions obtained at the epoch with the highest validation macro-F1 in each fold. Each of the 821 images contributes one prediction.

\begin{table}[!htbp]
    \caption{Transfer-learning comparison using ResNet34 models with affine classifier heads. Values are the mean $\pm$ sample standard deviation of pooled OOF metrics across three seeds.}
    \label{tab:transfer-results-full}
    \centering
    \small
    \setlength{\tabcolsep}{3pt}
    \begin{tabular}{@{}llccccc@{}}
        \toprule
        \shortstack[l]{Backbone\\fine-tuning} & Initialisation & Accuracy & \shortstack{Balanced\\Accuracy} & Macro-Precision & Macro-F1 & Weighted-F1 \\
        \midrule
        Full & ImageNet & $0.935 \pm 0.004$ & $0.891 \pm 0.009$ & $0.861 \pm 0.008$ & $0.874 \pm 0.008$ & $0.937 \pm 0.003$ \\
        Full & MLL23-adapted & $0.933 \pm 0.004$ & $0.901 \pm 0.012$ & $0.869 \pm 0.009$ & $0.881 \pm 0.004$ & $0.936 \pm 0.001$ \\
        \addlinespace
        Layers 3-4 & ImageNet & $0.932 \pm 0.004$ & $0.895 \pm 0.004$ & $0.862 \pm 0.004$ & $0.875 \pm 0.004$ & $0.935 \pm 0.002$ \\
        Layers 3-4 & MLL23-adapted & $0.935 \pm 0.009$ & $0.889 \pm 0.010$ & $0.872 \pm 0.008$ & $0.879 \pm 0.008$ & $0.936 \pm 0.008$ \\
        \addlinespace
        Layer 4 only & ImageNet & $0.921 \pm 0.003$ & $0.866 \pm 0.020$ & $0.835 \pm 0.009$ & $0.847 \pm 0.009$ & $0.925 \pm 0.003$ \\
        Layer 4 only & MLL23-adapted & $0.929 \pm 0.003$ & $0.869 \pm 0.007$ & $0.849 \pm 0.003$ & $0.858 \pm 0.002$ & $0.930 \pm 0.003$ \\
        \addlinespace
        Frozen & ImageNet & $0.786 \pm 0.002$ & $0.717 \pm 0.007$ & $0.681 \pm 0.004$ & $0.685 \pm 0.003$ & $0.802 \pm 0.002$ \\
        Frozen & MLL23-adapted & $0.869 \pm 0.010$ & $0.831 \pm 0.025$ & $0.776 \pm 0.011$ & $0.786 \pm 0.016$ & $0.881 \pm 0.009$ \\
        \bottomrule
    \end{tabular}
\end{table}

\begin{table}[!htbp]
    \centering
    \caption{Class-specific performance with and without intermediate MLL23 adaptation for ResNet34 models with a frozen backbone and an affine classifier head.}
    \label{tab:transfer-class-results}
    \begingroup
    \small
    \renewcommand{\arraystretch}{1.1}
    \setlength{\tabcolsep}{4pt}
    \begin{tabular*}{\linewidth}{@{\extracolsep{\fill}}lrlccc@{}}
        \toprule
        Class & Support & Initialisation & Precision & Recall & F1 \\
        \midrule
        \multirow{2}{*}{BAS} & \multirow{2}{*}{18} & ImageNet & $0.229 \pm 0.011$ & $0.463 \pm 0.032$ & $0.307 \pm 0.015$ \\
        & & MLL23-adapted & $0.408 \pm 0.030$ & $0.741 \pm 0.085$ & $0.526 \pm 0.040$ \\
        \addlinespace[4pt]
        \multirow{2}{*}{EOS} & \multirow{2}{*}{43} & ImageNet & $0.819 \pm 0.013$ & $0.806 \pm 0.013$ & $0.812 \pm 0.012$ \\
        & & MLL23-adapted & $0.939 \pm 0.012$ & $0.829 \pm 0.027$ & $0.881 \pm 0.010$ \\
        \addlinespace[4pt]
        \multirow{2}{*}{KSC} & \multirow{2}{*}{158} & ImageNet & $0.941 \pm 0.034$ & $0.916 \pm 0.013$ & $0.928 \pm 0.010$ \\
        & & MLL23-adapted & $0.976 \pm 0.004$ & $0.960 \pm 0.013$ & $0.968 \pm 0.009$ \\
        \addlinespace[4pt]
        \multirow{2}{*}{LYA} & \multirow{2}{*}{30} & ImageNet & $0.230 \pm 0.015$ & $0.500 \pm 0.033$ & $0.314 \pm 0.007$ \\
        & & MLL23-adapted & $0.329 \pm 0.040$ & $0.689 \pm 0.051$ & $0.445 \pm 0.047$ \\
        \addlinespace[4pt]
        \multirow{2}{*}{LYT} & \multirow{2}{*}{224} & ImageNet & $0.819 \pm 0.022$ & $0.731 \pm 0.037$ & $0.772 \pm 0.018$ \\
        & & MLL23-adapted & $0.914 \pm 0.030$ & $0.815 \pm 0.033$ & $0.862 \pm 0.022$ \\
        \addlinespace[4pt]
        \multirow{2}{*}{MON} & \multirow{2}{*}{180} & ImageNet & $0.825 \pm 0.018$ & $0.781 \pm 0.022$ & $0.802 \pm 0.011$ \\
        & & MLL23-adapted & $0.922 \pm 0.017$ & $0.891 \pm 0.025$ & $0.906 \pm 0.005$ \\
        \addlinespace[4pt]
        \multirow{2}{*}{NGS} & \multirow{2}{*}{168} & ImageNet & $0.908 \pm 0.018$ & $0.821 \pm 0.021$ & $0.862 \pm 0.013$ \\
        & & MLL23-adapted & $0.945 \pm 0.004$ & $0.889 \pm 0.012$ & $0.916 \pm 0.008$ \\
        \bottomrule
    \end{tabular*}
    \endgroup

    \par\vspace{4pt}
    \begin{minipage}{\linewidth}
        \footnotesize
        \raggedright
        Values are the mean $\pm$ sample standard deviation across three seeds. For each seed, five-fold out-of-fold (OOF) predictions are pooled for all 821 images. Metrics are calculated from the raw counts in the resulting confusion matrix. Support denotes the number of images per class.
    \end{minipage}
\end{table}

\clearpage

\begin{table}[!htbp]
    \caption{Aggregate affine-hybrid comparison using the MLL23-adapted ResNet34 backbone with residual layers 3-4 fine-tuned. In the configuration labels, $B$ denotes the number of independent 12-qubit blocks and $L$ the number of variational layers per block. Values are the mean $\pm$ sample standard deviation of pooled OOF metrics across three seeds.}
    \label{tab:head-results-full}
    \centering
    \small
    \setlength{\tabcolsep}{4pt}
    \begin{tabular}{@{}lccccc@{}}
        \toprule
        Head & Accuracy & \shortstack{Balanced\\Accuracy} & Macro-Precision & Macro-F1 & Weighted-F1 \\
        \midrule
        Affine & $0.935 \pm 0.009$ & $0.889 \pm 0.010$ & $0.872 \pm 0.008$ & $0.879 \pm 0.008$ & $0.936 \pm 0.008$ \\
        Hybrid $B1/L1$ & $0.918 \pm 0.011$ & $0.876 \pm 0.008$ & $0.842 \pm 0.020$ & $0.854 \pm 0.013$ & $0.922 \pm 0.010$ \\
        Hybrid $B2/L1$ & $0.924 \pm 0.009$ & $0.886 \pm 0.008$ & $0.850 \pm 0.010$ & $0.865 \pm 0.010$ & $0.928 \pm 0.008$ \\
        \bottomrule
    \end{tabular}
\end{table}

\begin{table}[!htbp]
    \caption{Pooled OOF precision, recall, and F1 for the three least-supported classes: basophils (BAS, $n=18$), atypical lymphocytes (LYA, $n=30$), and eosinophils (EOS, $n=43$). All pipelines use the MLL23-adapted ResNet34 backbone with residual layers 3-4 fine-tuned. Both hybrid heads use $L=1$, with independently simulated 12-qubit blocks. Values are the mean $\pm$ sample standard deviation across three seeds.}
    \label{tab:low-support-results}
    \centering
    \small
    \setlength{\tabcolsep}{5pt}
    \begin{tabular}{@{}llccc@{}}
        \toprule
        Class & Head & Precision & Recall & F1 \\
        \midrule
        \multirow{3}{*}{BAS} & Affine & $0.798 \pm 0.041$ & $0.870 \pm 0.032$ & $0.832 \pm 0.029$ \\
        & Hybrid $B1/L1$ & $0.710 \pm 0.103$ & $0.852 \pm 0.085$ & $0.768 \pm 0.047$ \\
        & Hybrid $B2/L1$ & $0.715 \pm 0.043$ & $0.870 \pm 0.032$ & $0.784 \pm 0.020$ \\
        \addlinespace
        \multirow{3}{*}{LYA} & Affine & $0.478 \pm 0.076$ & $0.567 \pm 0.058$ & $0.517 \pm 0.063$ \\
        & Hybrid $B1/L1$ & $0.385 \pm 0.070$ & $0.567 \pm 0.100$ & $0.458 \pm 0.081$ \\
        & Hybrid $B2/L1$ & $0.423 \pm 0.070$ & $0.589 \pm 0.069$ & $0.490 \pm 0.060$ \\
        \addlinespace
        \multirow{3}{*}{EOS} & Affine & $0.977 \pm 0.000$ & $0.977 \pm 0.000$ & $0.977 \pm 0.000$ \\
        & Hybrid $B1/L1$ & $0.970 \pm 0.026$ & $0.969 \pm 0.013$ & $0.969 \pm 0.006$ \\
        & Hybrid $B2/L1$ & $0.970 \pm 0.013$ & $0.984 \pm 0.013$ & $0.977 \pm 0.011$ \\
        \bottomrule
    \end{tabular}
\end{table}

\FloatBarrier

\section{Detailed training costs}
\label{sec:supp-costs}

\begin{table}[!htbp]
    \caption{Observed end-to-end training runtimes and trainable head parameter counts for the affine baseline and seven hybrid configurations.  Each runtime is the total time recorded in Weights \& Biases across the five cross-validation folds for seed $250$. All configurations are trained on a single NVIDIA A40 GPU for 50 epochs per fold. Hybrid circuits are evaluated by noiseless statevector simulation.}
    \label{tab:computational-cost}
    \centering
    \small
    \begin{tabular}{lrr}
        \toprule
        Configuration & Trainable head params. & CV runtime (h) \\
        \midrule
        Affine & $3\ 591$ & 0.224 \\
        \addlinespace
        Single encoding $B1/L1$ & $19\ 943$ & 22.716 \\
        Single encoding $B2/L1$ & $38\ 855$ & 44.333 \\
        Single encoding $B3/L1$ & $57\ 767$ & 66.698 \\
        \addlinespace
        Standard re-uploading $B1/L2$ & $20\ 027$ & 34.613 \\
        \addlinespace
        Incremental re-uploading $B1/L2$ & $38\ 495$ & 34.519 \\
        Incremental re-uploading $B1/L3$ & $57\ 047$ & 47.172 \\
        Incremental re-uploading $B2/L2$ & $75\ 959$ & 70.354 \\
        \bottomrule
    \end{tabular}
\end{table}

\clearpage

\section{Seed-specific confusion matrices}
\label{sec:supp-confusion}

\begin{figure}[!htbp]
    \centering
    \includegraphics[width=\linewidth]{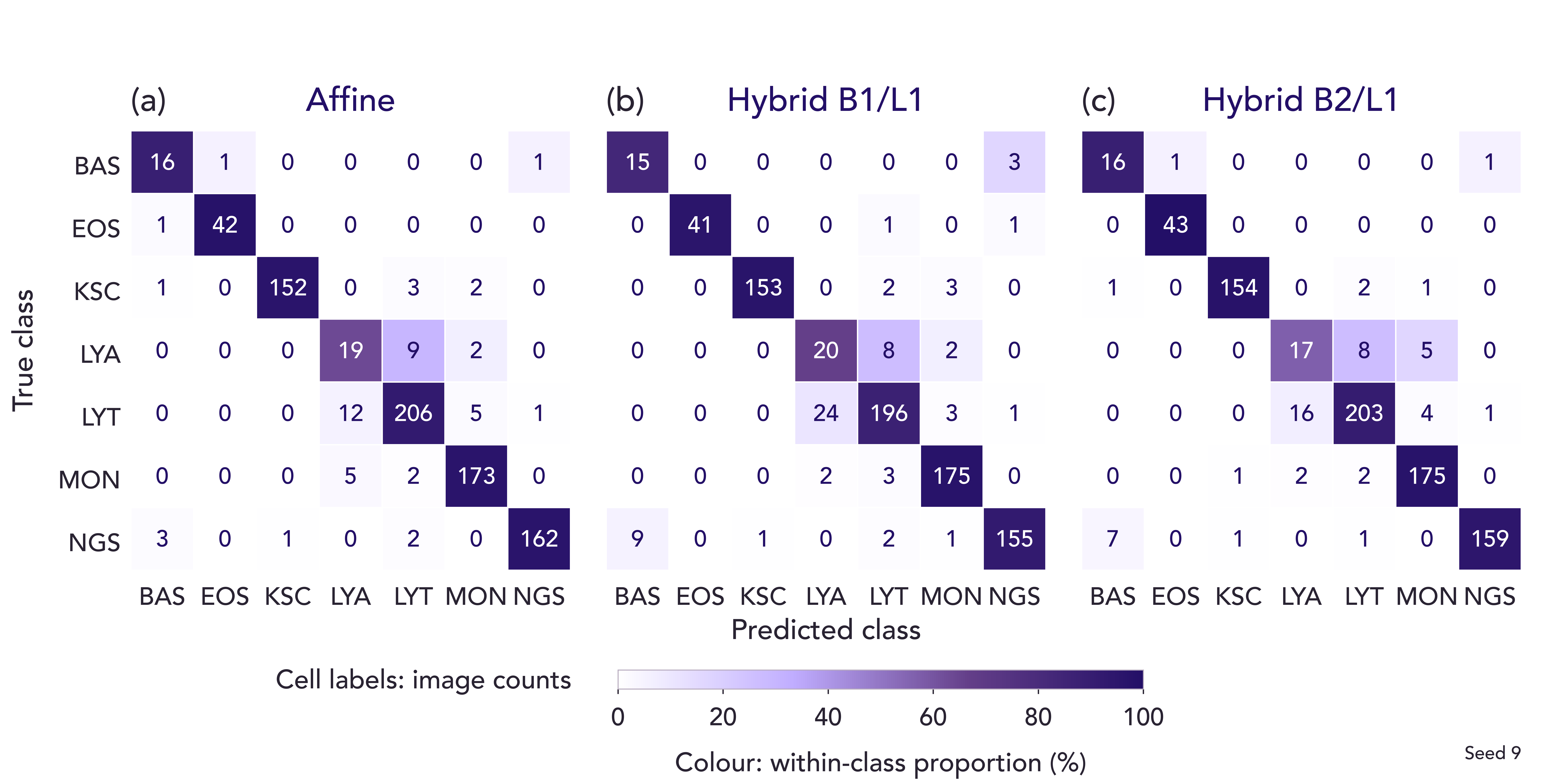}
    \caption{Pooled OOF confusion matrices for seed 9. From left to right: Affine, Hybrid $B1/L1$, and Hybrid $B2/L1$. Cell annotations indicate image counts, whereas colour represents percentages within each true class}
    \label{fig:confusion-seed9}
\end{figure}

\begin{figure}[!htbp]
    \centering
    \includegraphics[width=\linewidth]{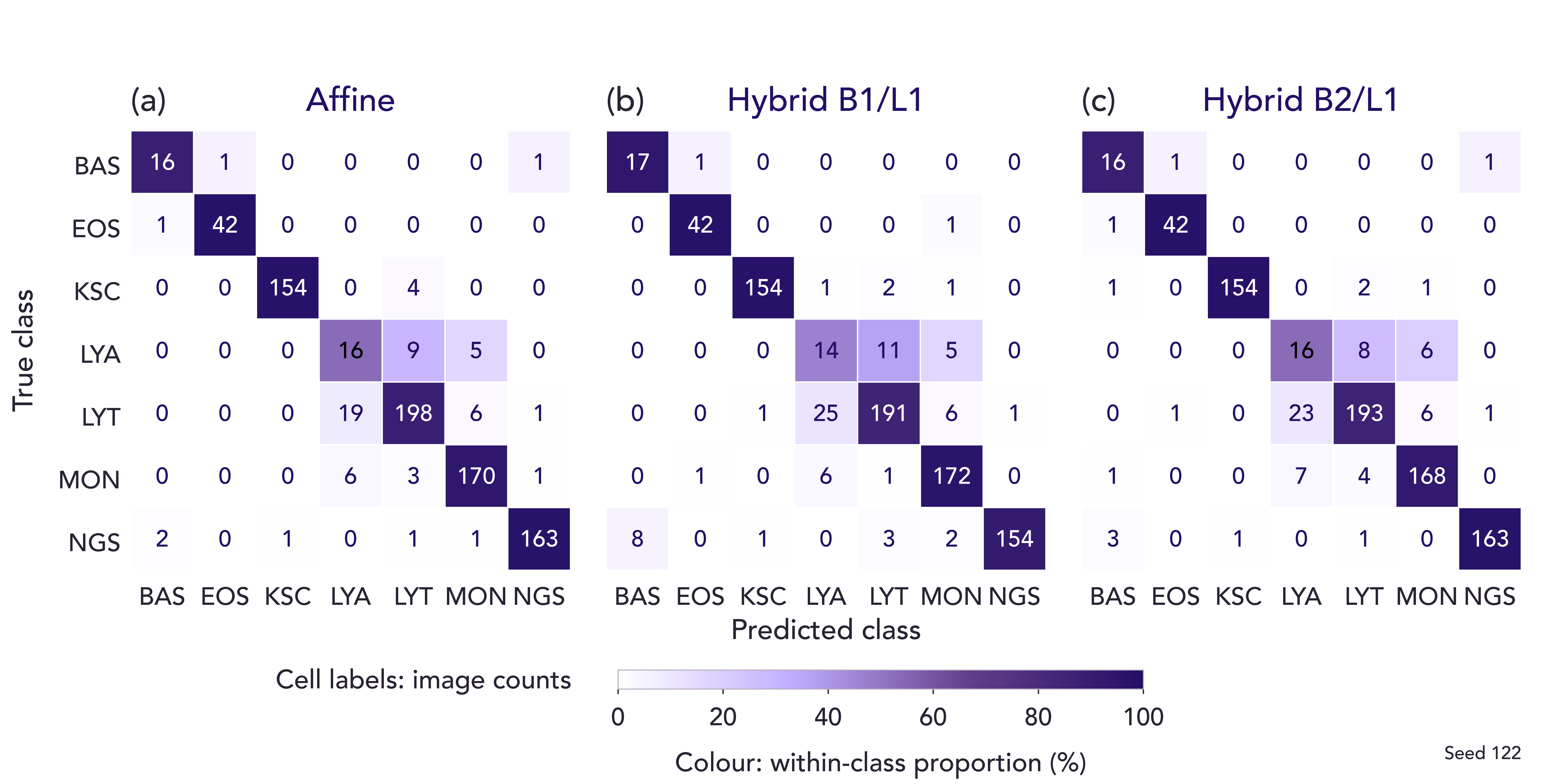}
    \caption{Pooled OOF confusion matrices for seed 122. From left to right: Affine, Hybrid $B1/L1$, and Hybrid $B2/L1$. Cell annotations indicate image counts, whereas colour represents percentages within each true class}
    \label{fig:confusion-seed122}
\end{figure}

\begin{figure}[!htbp]
    \centering
    \includegraphics[width=\linewidth]{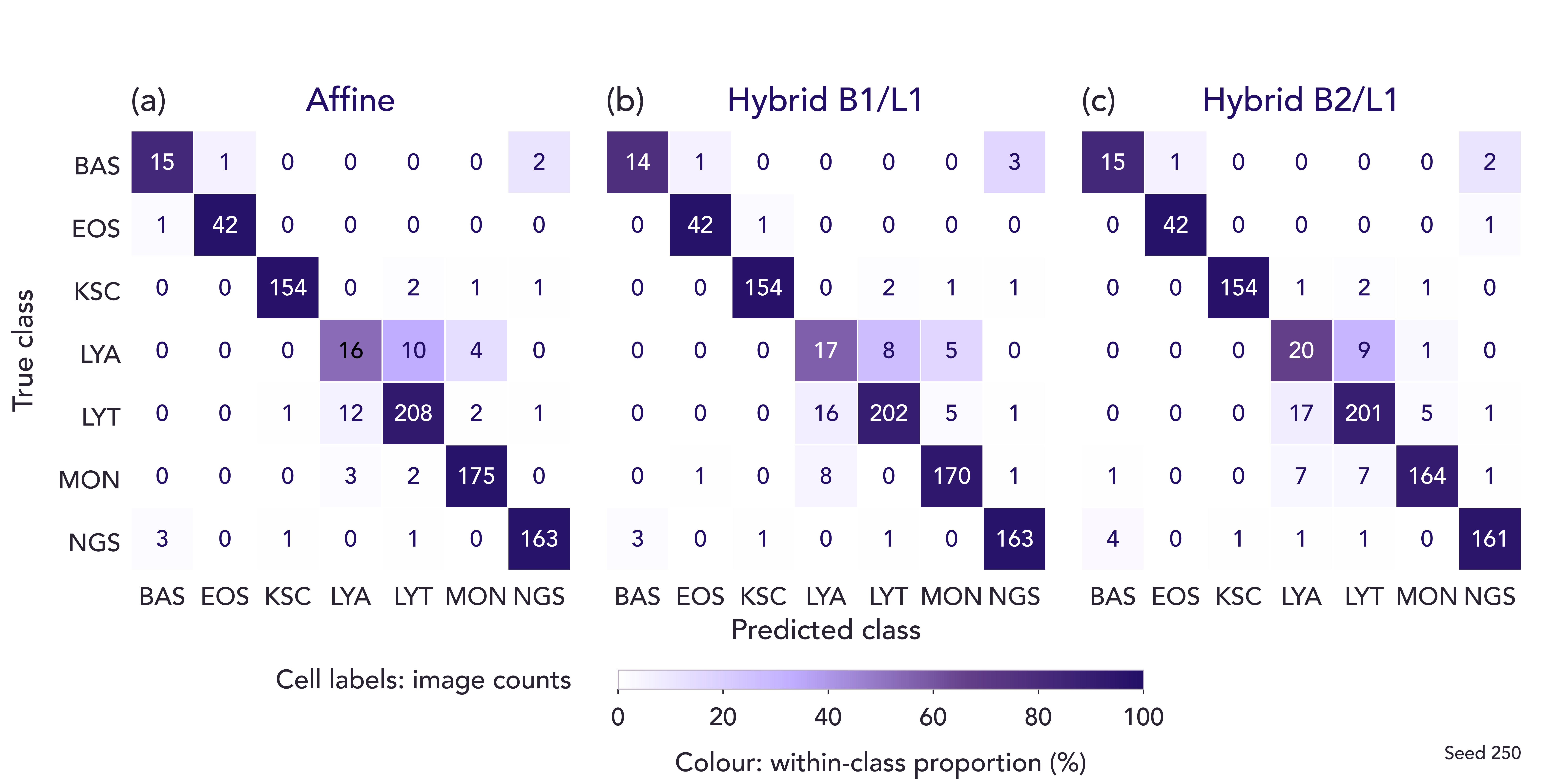}
    \caption{Pooled OOF confusion matrices for seed 250. From left to right: Affine, Hybrid $B1/L1$, and Hybrid $B2/L1$. Cell annotations indicate image counts, whereas colour represents percentages within each true class}
    \label{fig:confusion-seed250}
\end{figure}

Figures~\ref{fig:confusion-seed9}-\ref{fig:confusion-seed250} present the pooled OOF confusion matrices for each seed and classifier head. All three pipelines use the MLL23-adapted ResNet34 backbone with residual layers 3-4 fine-tuned.

Rows represent true classes and columns represent predicted classes.

\clearpage

\end{document}